\documentclass[prd, floatfix,
 reprint,
superscriptaddress,
groupedaddress,
unsortedaddress,
nofootinbib,
 amsmath,amssymb,
 aps,
]{revtex4-2}
\usepackage{amsmath}
\usepackage{amssymb}

\usepackage{graphicx}
\usepackage[dvipsnames]{xcolor}
\definecolor{light gray}{gray}{0.8}
\usepackage{pgfplots}
\usepackage{tikz}

\usetikzlibrary{decorations.markings} 

\usepackage{placeins}
\usepackage{dcolumn}
\usepackage{bm}
\definecolor{prdlink}{RGB}{0,0,128}\usepackage[colorlinks=true,linkcolor=prdlink,citecolor=prdlink,urlcolor=prdlink]{hyperref}
\usepackage{orcidlink}

\pgfplotsset{compat=1.18}\begin{document}

\title{Lensing in matched exterior and interior Kottler solutions}

\author{Mourad Guenouche\,\orcidlink{0000-0002-1211-4940}}
\email{guenouche\_mourad@umc.edu.dz}\email{guenouche\_mourad@univ-khenchela.dz}
  
\affiliation{Laboratoire de Physique Théorique, \href{https://ror.org/017wv6808}{Université Frères Mentouri-Constantine 1}, BP 325 route de Ain El Bey, 25017 Constantine, Algeria \\
and Département des Sciences de la Matière, \href{https://ror.org/02yyskm09}{Université Abbès Laghrour-Khenchela}, BP 1252, El Houria, Route de Constantine, 40004 Khenchela, Algeria}

\date{\today}

\begin{abstract}
We analyze gravitational lensing in a matched Kottler spacetime with cosmological constant $\Lambda$, where a uniform-density fluid sphere described by the interior Kottler solution is smoothly joined to the exterior Kottler region. We derive an analytic expression for the deflection angle of light rays traversing both regions, isolating the corrections induced by the interior mass distribution relative to the vacuum Kottler case. The interior terms systematically reduce the bending angle and tangential magnification while enhancing the radial one, leading to a scale-dependent effect on the total magnification. Numerical evaluation shows a mild amplification for galaxy lenses and a demagnification for clusters, reflecting the contrast between compact and diffuse configurations. The induced $\Lambda$-contribution from the interior corrections further reduces the bending, but only slightly, consistent with its role in the vacuum Kottler case. These results demonstrate that lensing observables retain a measurable imprint of interior structure in SdS-like lenses, crucial for refining realistic strong-lensing models.
\end{abstract}

\maketitle
\section{Introduction}
Gravitational lensing is a sensitive probe of both local matter distributions and the geometry of the Universe. While the Schwarzschild solution describes deflection by point-like masses, realistic astrophysical lenses such as galaxies and clusters possess extended interiors and are embedded in a cosmological background. The Kottler or Schwarzschild–de Sitter (SdS) solution incorporates the cosmological constant $\Lambda$, whose impact on lensing has been actively debated since Rindler and Ishak \cite{Rindler2007}. In the exterior SdS spacetime, $\Lambda$ affects measurable angles only through geometry, with negligible effect compared to mass. In interior SdS extensions, however, $\Lambda$ explicitly enters the geodesic equations due to the nonvacuum matter profile, raising the question of how both the cosmological constant and finite-size structure alter standard predictions \cite{Schücker2010}.

Cluster-lensed quasars provide a natural setting for this problem. Some images form very close to or even inside the lens mass distribution, corresponding to photon trajectories that traverse a fluid or cored interior. Such images, typically faint and demagnified, motivate extending lens models beyond the exterior SdS case \cite{Joshua,Inada,Ertl}.

Despite progress on the roles of $\Lambda$ and interior matter in lensing, a unified treatment of lensing in matched SdS spacetime—where an extended interior is smoothly joined to the exterior Kottler region—remains of interest for bridging the gap between idealized point-mass lens models and realistic extended astrophysical systems and for assessing the extent to which lensing observables encode information about both cosmological curvature and interior structure. 

\section{Matched Kottler metric and light deflection angle}

We model the lens as a uniform-density fluid sphere, of total mass $M$ and radius $r_{\rm B}$, matched smoothly to the exterior SdS solution using the exact global metric
\begin{equation}
{\rm d}s^2=f(r){\rm d}t^2-g(r)^{-1}{\rm d}r^2-r^2\left({\rm d}\theta^2+\sin^2\theta{\rm d}\varphi^2\right),\label{SdSint}
\end{equation}
where $f(r)=g(r)=1-\delta(r)-\lambda(r)^{2}$, for $r\geq r_{{\rm B}}$, and
\begin{align}
 f(r)&=\frac{1}{4}\left[ \frac{3\delta _{{\rm B}}}{\delta _{{\rm B}}+\lambda_{{\rm B}}^{2}}\sqrt{g(r_{{\rm B}})}
-\left(\frac{3\delta _{{\rm B}}}{\delta _{{\rm B}}+\lambda_{{\rm B}}^{2}}-2\right) \sqrt{g(r)}
\right] ^{2},\nonumber\\
g(r)&=1-\left(\delta _{{\rm B}}+\lambda_{{\rm B}}^{2}\right)\frac{r^{2}}{r_{{\rm B}}^{2}},
\end{align}
for $r\leq r_{{\rm B}}$, with $\delta (r)=2GM/r$, $\lambda (r)=\sqrt{\Lambda /3}r$, $\delta_{\rm B}=\delta(r_{\rm B})$, and $\lambda_{{\rm B}}=\lambda(r_{{\rm B}})$.

The photon trajectory lies in the equatorial plane $\theta=\pi/2$, with azimuthal variation ruled by the equation
\begin{equation}
{\rm d}\varphi (r)=\pm \frac{{\rm d}r}{u(r)},\quad u(r)=r\sqrt{
g(r) }\left( \dfrac{f(r_{0})r^{2}}{r_{0}^{2}f(r)}-1\right)^{\frac{1}{2}}, \label{defeqt}
\end{equation}
where $r_0$ is the pericenter inside the SdS fluid $(r_{0}\leq r_{{\rm B}})$ (\ref{def}), related to the conserved angular momentum per unit mass, associated with the $\varphi$-component $J(=r^2\dot{\varphi})$, by $J=\pm r_0/\sqrt{f(r_0)}$, The second constant of motion associated with the $t$-component is absorbed into a redefinition of the affine parameter $(E=1)$. For configurations where $\varphi$ increases along the path, it is natural to take $J>0$, while preserving full generality by $J\to -J$. The plus and minus signs in Eq.~(\ref{defeqt}) reflect outgoing and ingoing geodesics relative to the lens. The reception angle $\alpha$ at Earth $(r_{\rm E},\pi)$ $(r_{\rm E}\gg r_{{\rm B}})$ and the emission angle $\sigma$ at the source $(r_{\rm S},\varphi_{\rm S})$ $(r_{\rm S}\gg r_{{\rm B}})$ relate to $r_0$ by the tetrad projection, 
\begin{equation}
J(r_0)=\frac{r_{\rm E}\sin\alpha}{\sqrt{f(r_{\rm E})}}=\frac{r_{\rm S}\sin\sigma}{\sqrt{f(r_{\rm S})}}.
\label{r0rel}
\end{equation}
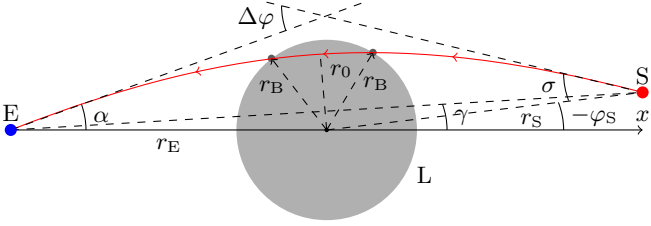
\begin{figure}[ht]
\centering
\begin{tikzpicture}
 
\fill[gray!62] (0,0) circle [radius=1.2cm]; 
\node at (1.3,-0.6) {L};
 
\draw[->, thin ] (-4.2,0) -- (4.2,0) node[pos=0.25, below] {$r_{\rm E}$} node[pos=1, above]{$x$}; 

\draw[dashed](-4.2,0) -- (4.2,0.5);
\draw[dashed](0,0) -- (4.2,0.5)  node[pos=0.65, below] {$r_{\rm S}$};

\fill[darkgray!85] (0.615,1.03) circle [radius=0.05cm];
\draw[dashed, ->, thin](0,0) -- (0.615,1.03) node[pos=0.6, right]{$r_{\rm B}$};

\fill[darkgray!85] (-0.734,0.952) circle [radius=0.05cm];
\draw[dashed, ->, thin](0,0) -- (-0.734,0.952) node[pos=0.6, left]{$r_{\rm B}$};

\draw[dashed, thin](0,0) -- (-0.09,1.05) node[pos=0.75, right]{$r_{0}$};

\draw[thin, black] (1.6,0) arc (0:20:1cm) node[pos=0.5, right]{$\gamma$};
\draw[thin, black] (3.15,0) arc (0:21.5:1cm) node[pos=0.5, right]{$-\varphi_\text{S}$};
\draw[thin, black] (-3.2,0) arc (0:20.35:1cm) node[pos=0.5, right]{$\alpha$};

\coordinate (A) at (4.2,0.5);
\coordinate (C) at (-4.2,0);
\draw[color=red, postaction={ decorate, decoration={ markings, mark=at position 0.3 with {\arrow{>}}, mark=at position 0.5 with {\arrow{>}}, mark=at position 0.7with {\arrow{>}}}}] (A) to [bend left=-18] (C);

\draw[dashed](-4.2,0) -- (0.5,1.7);
\draw[dashed](-0.8,1.7) -- (4.2,0.5);
\draw[thin, black] (-0.6,1.656) arc (0:19.8:-1cm) node[pos=0.5, left]{$\Delta\varphi$};

\draw[thin, black] (3.15,0.75) arc (0:21:-1cm) node[pos=0.5, left]{$\sigma$};

\fill[red] (4.2,0.5) circle [radius=0.08cm] node[above, text=black]{S};
\fill[blue] (-4.2,0) circle [radius=0.08cm] node[above, text=black]{E};
\fill[black] (0,0) circle [radius=0.03cm];

\end{tikzpicture} 
\caption{A light ray emitted by a source S, bent outside and inside the SdS fluid lens L, and received at the blue Earth E.}
\label{def}
\end{figure}
The deflection angle is $\Delta\varphi=\alpha+\sigma-\varphi_{\rm S}$, where $\varphi_{\rm S}$ is obtained by integrating Eq.~(\ref{defeqt}) across both regions,
\begin{equation}
\varphi_{\rm S}=\pi -\left( \int_{r_{{\rm B}}}^{r_{{\rm E}}}+\int_{r_{
{\rm B}}}^{r_{{\rm S}}}\right) \frac{{\rm d}r}{u_{{\rm ex}}(r)}
-2\int_{r_{0}}^{r_{{\rm B}}}\frac{{\rm d}r}{u_{{\rm in}}(r)},
\label{def}
\end{equation}
where $u_{\text{ext}}(r)$ and $u_{\text{int}}(r)$ are the relevant expressions of $u(r)$ (\ref{defeqt}) for the exterior and interior SdS metrics. These integrals are denoted by $I_{\rm E}$, $I_{\rm S}$ and $I_{\rm B}$, respectively, and evaluated to first order in $\delta_{{\rm B}}=2GM/r_{{\rm B}} \ll 1$, as appropriate for non-compact bodies whose sizes greatly exceed their Schwarzschild radii. We extend this for photons crossing into the interior SdS region, in the sense that approximating in terms of $\delta_0=\delta(r_0)=2GM/r_0$ amounts to the same thing, provided $r_{{\rm Schw}}\ll r_0\leq r_{\rm B}$. We obtain
\begin{align}
I_{\rm E}&\simeq\arcsin \frac{r_{0}}{r_{{\rm B}}}-\arcsin \frac{r_{0}}{r_{{\rm E}}}\nonumber\\
&+\frac{\delta _{0}}{2}\left\{C(r_{{\rm E}})-C(r_{{\rm B}})+(1-h_0)[\varepsilon(r_{\rm E})-\varepsilon(r_{\rm B})]\right\},\label{int_uext}\\
I_{\rm B}&\simeq \frac{\pi }{2}-\arcsin \frac{r_{0}}{r_{{\rm B}}}\nonumber\\
&+\frac{\delta _{0}}{2}\left[ C(r_{{\rm B}})+(1-h_0)\varepsilon(r_{\rm B})-2\left( 1 -\frac{r_{0}^{2}}{r_{{\rm B}}^{2}}\right)^{\frac{3}{2}}\right],\label{int_uint2}
\end{align}
where $C(r_{{\rm E}})$, $C(r_{{\rm B}})$, $\varepsilon(r_{\rm E})$, and $\varepsilon(r_{\rm B})$ are defined by
\begin{align}
C(r)&=\left(1-\frac{r_{0}^{2}}{r^{2}}\right)^{\frac{1}{2}}\left[1+\left(1+\frac{r_{0}}{r}\right)^{-1}\right],\label{C_function}\\\varepsilon(r) &=\frac{r_{0}}{r}\left( 1-\frac{r_{0}^{2}}{r^{2}}\right) ^{-\frac{1}{2}}\label{varepsilon_function}.
\end{align}
Here, $h_0=h(r_0)$ $\left( r\leq r_{{\rm B}}\right)$ is given by
\begin{equation}
h(r)=\dfrac{r^3}{r_{{\rm B}}^3}\left( 3\frac{1-\sqrt{1-\lambda _{{\rm B}}^{2}}\sqrt{1-\lambda (r)^{2}}}{\lambda (r)^{2}}-2 \right),
\end{equation}
which characterizes the first order correction to $f(r)$,
\begin{equation}
f(r)\simeq 1-\delta (r)h(r)-\lambda (r)^{2}\quad \left( r\leq r_{{\rm B}}\right),\label{f_approx}
\end{equation}
with $0<h(r)\leq h(r_{\rm B})=1$, ensuring continuity at the boundary.
A similar formula to (\ref{int_uext}) holds for $I_{\rm S}$, replacing $r_{\rm E}$ by $r_{\rm S}$. Inserting $I_{\rm E}$, $I_{\rm B}$, and $I_{\rm S}$ back into~(\ref{def}), the term $\delta_0[C(r_{{\rm B}})+(1-h_0)\varepsilon(r_{\rm B})]$ drops out, yielding
\begin{equation}
\Delta\varphi \simeq  \Delta\varphi_{\rm ex}-2\delta_0\left[ \left(1-\frac{r_0^2}{r_{\rm B}^2}\right)^{\frac{3}{2}}-(1-h_0)\frac{\varepsilon(r_{\rm E},r_{\rm S})}{4}\right]\label{total_def},
\end{equation}
where $\varepsilon(r_{\rm E},r_{\rm S})=\varepsilon(r_{\rm E})+\varepsilon(r_{\rm S})$. The $\delta_0$-term represents a subtractive correction induced by the SdS fluid, which vanishes for a grazing trajectory, $r_{\rm B}=r_0$ $(h_0=1)$, recovering the standard deflection in the vacuum SdS,
\begin{align}
\Delta\varphi_{\rm ex}&\simeq \alpha+\sigma-\arcsin\frac{r_0}{r_{\rm E}}-\arcsin\frac{r_0}{r_{\rm S}}\nonumber\\& +2\delta_0\left[ C(r_{\rm E})+C(r_{\rm S})\right].
\end{align}
The light deflection is explicitly offset by $2\delta_0(1-r_0^2/r_{\rm B}^2)^{3/2}$ as long as $r_0<r_{\rm B} $, whereas it is  slightly enhanced by the term $\delta_0(1-h_0)\varepsilon(r_{\rm E},r_{\rm S})/2$, which depends on small ratios $r_0/r_{\rm E}$ and $r_0/r_{\rm S}$ (\ref{varepsilon_function}), with the Earth and source being very far from the lens for typical situations. Physically, the mass enclosed up to the radius, growing as $r^3$ for a uniform density, reflects a weaker curvature inside the fluid sphere than outside, where the mass is fully located within that radius. Specifically, this additive term includes, through $h_0$, a $\Lambda$ contribution that reduces the light bending, consistent with its effect outside the mass distribution \cite{Rindler2007}. This reduction is very small and can be made explicit by expanding $h_{0}$ to first order in $\lambda
_{{\rm B}}$,
\begin{equation}
h_{0}\simeq \frac{1}{2}\frac{r_{0}}{r_{{\rm B}}}\left[ 3-\frac{r_{0}^{2}}{
r_{{\rm B}}^{2}}+\left( 1-\frac{r_{0}^{2}}{r_{{\rm B}}^{2}}\right) ^{2}
\frac{\Lambda }{4}r_{{\rm B}}^{2}\right],\label{h_approx}
\end{equation}
then substituting into (\ref{total_def}),
\begin{align}
\Delta\varphi &\simeq  \Delta\varphi_{\rm ex}-2\delta_0\left[ \left(1-\frac{r_0^2}{r_{\rm B}^2}\right)^{\frac{3}{2}}-\frac{\varepsilon(r_{\rm E},r_{\rm S})}{4}\left( 1+\frac{1}{2}\frac{r_{0}}{r_{{\rm B}}}\right)\right.\nonumber\\ 
&\left.\times\left( 1-\frac{r_{0}}{r_{{\rm B}}}\right)^{2}\right]-\delta_{\rm B}\frac{\varepsilon(r_{\rm E},r_{\rm S})}{4}\left( 1-\frac{r_{0}^{2}}{r_{{\rm B}}^{2}}\right)^{2}\frac{\Lambda }{4}r_{{\rm B}}^{2}\label{total_def_Lambda}.
\end{align} 
This remains true if one considers only the bending within the SdS fluid. Substituting (\ref{h_approx}) into (\ref{int_uint2}), one gets
\begin{align}
2I_{{\rm B}}-\pi&\simeq  -2\arcsin \frac{r_{0}}{r_{{\rm B}}}+3\frac{GM}{r_{\rm B}}
\frac{r_{0}}{r_{{\rm B}}}\left( 1-\frac{r_{0}^{2}}{r_{{\rm B}}^{2}}
\right) ^{\frac{1}{2}}\nonumber\\
&-\frac{GM}{r_0}\left( 1-\frac{r_{0}^{2}}{r_{{\rm B}}^{2}}
\right) ^{\frac{3}{2}}\frac{\Lambda }{4}r_{0}^{2}.\label{int_Lambda}
\end{align}
In Ref.~\cite{Schücker2010}, Schücker confirmed this direct dependence of light bending on $\Lambda$ for a light ray confined inside a global SdS fluid--unlike the exterior case--but found that the $\Lambda$-effect appears reversed compared to the usual expectation. As stated earlier, the subtractive $\Lambda$-term in (\ref{int_Lambda}) is, however, suppressed by an additive one arising from $I_{\rm E}$ and $I_{\rm S}$ (\ref{int_uext}), leaving only residual dependencies encoded in $r_0$ via $2\delta_0(1-r_0^2/r_{\rm B}^2)^{3/2}$ (\ref{int_uint2}), subsequently added, along with a remaining $\Lambda$-term~(\ref{total_def_Lambda}), to the total bending. Furthermore, the bending inside the fluid increases linearly with the compactness parameter $\delta_{\rm B}$, consistent with physical intuition that a more compact configuration induces stronger deflection. Although both corrections stem from the same central mass, the subtractive $\delta_0$-term (\ref{total_def}), $2\delta_0(1-r_0^2/r_{\rm B}^2)^{3/2}$, in the total bending accounts, as noted, for the reduced interior curvature relative to the vacuum case, while the additive $\delta_{\rm B}$-term (\ref{int_Lambda}) in the interior bending reflects the cumulative effect of the distributed mass along the light path. Using (\ref{f_approx}) with (\ref{h_approx}) in Eq.~(\ref{r0rel}) leads to, after a redefinition of the impact parameter, a quadratic,
\begin{equation}
a\frac{r_{0}^{4}}{r_{{\rm B}}^{4}}+\left( \frac{r_{{\rm B}}^{2}}{J_{\Lambda}^{2}}-b\right) \frac{r_{0}^{2}}{r_{{\rm B}}^{2}}+c=1,\quad \frac{1}{J_{\Lambda}^{2}}=\frac{1}{J^{2}}+\frac{\Lambda}{3}\label{quad_eq}
\end{equation}
solved by
\begin{equation}
r_{0}=\frac{r_{{\rm B}}}{\sqrt{2a}}\left[{\sqrt{\left( \dfrac{r_{{\rm B}}^{2}}{J_{\Lambda}^{2}}-b\right) ^{2}+4a\left( 1-c\right) }-\dfrac{r_{{\rm B}}^{2}}{J_{\Lambda}^{2}}+b}\right]^{\frac{1}{2}} ,\label{root0}
\end{equation}
where $a$, $b$ and $c$ are small coefficients defined by
\begin{equation}
a=\frac{3}{4}\frac{\delta _{{\rm B}}}{2}\lambda _{{\rm B}}^{2},\quad b=
\frac{\delta _{{\rm B}}}{2} +2a,\quad c=3\frac{\delta _{{\rm B}}}{2}+a.
\end{equation}
This generalizes the pericenter expression to account for a trajectory that penetrates the SdS fluid. The grazing case $(r_{\rm B}=r_0)$ reduces Eq.~(\ref{quad_eq}) to a depressed cubic, whose real root determines $r_0$ in the vacuum SdS \cite{Finelli2007,Ishak2010}. Linearizing (\ref{root0}) to leading order in $\delta_{\rm B}$ and $\lambda_{\rm B}$ yields
\begin{align}
r_{0}&\simeq J_{\Lambda}\left[ 1-\frac{1}{2}\frac{\delta_{{\rm B}}}{2}\left( 3-
\frac{J_{\Lambda}^{2}}{r_{{\rm B}}^{2}}\right)-\frac{3}{2}\frac{\delta_{{\rm B}}^2}{4}\frac{J_{\Lambda }^{2}}{r_{{\rm B}}^{2}} \right.\nonumber\\
&\left.-\frac{3 }{8}\frac{\delta_{{\rm B}}}{2}\left( 1-2\frac{J_{\Lambda }^{2}}{r_{{\rm B}}^{2}}\right)\lambda_{{\rm B}}^{2}\right],\\
&\simeq J\left[ 1-\frac{1}{2}\frac{GM}{r_{{\rm B}}}\left( 3-\frac{J^{2}}{r_{{\rm B}}^{2}}\right) -\frac{\Lambda }{6}J^{2}\right],
\end{align}
which shows that $r_0$ decreases linearly with both  $\delta_{\rm B}$ and $\Lambda$. The first reflects the focusing effect of increasing interior mass, while the latter, though counterintuitive, originates from the geometric dilution of spacetime curvature by $\Lambda$, which weakens the gravitational attraction. Although the light trajectory approaches more closely to the center, the overall bending is reduced—consistent with the effect observed in the vacuum SdS~\cite{Rindler2007,Sereno2008}, where the $\Lambda$-term remains identical even for penetrating trajectories. Sereno~\cite{Sereno2008} confirms this effect by showing that $\Lambda$ reduces the Einstein-ring radius and leads to a slight demagnification of lensing images, consistent with a reduced deflection relative to the pure Schwarzschild.

Given that the Earth and the source are located at large but finite distances from the lens, $2GM/r_{{\rm E}}\ll \lambda_{{\rm E}}^2 (=\Lambda r_{{\rm E}}^2/3)$ and $2GM/r_{{\rm S}}\ll \lambda_{{\rm S}}^2(=\Lambda r_{{\rm S}}^2/3)$, the angles are small enough justifying these approximations of (\ref{r0rel})
\begin{equation}
r_{0}\simeq J\simeq\frac{\alpha r_{{\rm E}}}{\sqrt{1-\lambda_{{\rm E}}^{2}}}\simeq\frac{\sigma r_{{\rm S}}}{\sqrt{1-\lambda_{{\rm S}}^{2}}}\label{small_angles}.
\end{equation}
This allows the bending angle to be expressed in terms of measurable quantities as
\begin{align}
\Delta \varphi&\simeq \Delta \varphi _{{\rm ex}}-\frac{4GM}{\alpha r_{{\rm E}}}\sqrt{1-\lambda_{{\rm E}}^{2}}\left( 1-\frac{\alpha^{2}}{\beta^{2}}\right) ^{\frac{3}{2}},\label{deftotal}\\
&\simeq \frac{4GM}{\alpha r_{{\rm E}}}\sqrt{1-\lambda_
{{\rm E}}^{2}}\left[1-\left( 1-\frac{\alpha^{2}}{\beta^{2}}\right)^{\frac{3}{2}}\right]-\alpha\eta(\Lambda),\label{ext-int_bending}
\end{align}
where the angle $\beta=(r_{\rm B}/r_{\rm E})\sqrt{1-\lambda_{{\rm E}}^{2}}$ is introduced by analogy with (\ref{small_angles}), and
\begin{equation}
\Delta \varphi _{{\rm ex}}\simeq \frac{4GM}{\alpha r_{{\rm E}}}\sqrt{1-\lambda_
{{\rm E}}^{2}}-\alpha\eta(\Lambda),\label{ext_bending}
\end{equation}
where $\eta(\Lambda)$ is a correction term arising from the presence of $\Lambda$, defined by
\begin{equation}
\eta(\Lambda)=\frac{1-\sqrt{1-\lambda_{{\rm E}}^{2}}}{\sqrt{1-\lambda_{{\rm E}}^{2}}}+\frac{r_{{\rm E}}}{r_{{\rm S}}}\frac{1-\sqrt{1-\lambda_{{\rm S}}^{2}}}{\sqrt{1-\lambda_{{\rm E}}^{2}}}\label{eta}.
\end{equation}
For $\Lambda=0$ $[\eta=0]$, (\ref{ext-int_bending}) recovers the matched exterior-interior Schwarzschild bending angle, where (\ref{ext_bending}) reduces to the standard Einstein (or Schwarzschild) bending angle.
\section{Lens equation and magnification}
The interior fluid sphere structure modifies the lens equation $\gamma=\alpha-(d_{\rm LS}/d_{\rm S})\Delta \varphi$ \cite{Schneider1992}, relating the bending angle and the unlensed angular position of the source $\gamma$,
\begin{equation}
\gamma=\alpha\left[1+\frac{d_{\rm LS}}{d_{\rm S}}\eta(\Lambda)\right]-\frac{4GM}{\alpha d_{\rm L}}\frac{d_{\rm LS}}{d_{\rm S}}\left[1-\left( 1-\frac{\alpha^{2}}{\beta^{2}}\right)^{\frac{3}{2}}\right],
\end{equation}
by using angular diameter distances, $d_{\rm L}$ (Earth-to-lens), $d_{\rm S}$ (Earth-to-source), and $d_{\rm LS}$ (lens-to-source), derived in the de Sitter geometry à la Sereno \cite{Sereno2008},
\begin{equation}
d_{\rm L}=\frac{r_{\rm E}}{\sqrt{1-\lambda_{{\rm E}}}},\quad d_{\rm S}=\frac{r_{\rm E}+r_{\rm S}}{\sqrt{1-\lambda_{{\rm E}}}},\quad d_{\rm LS}=r_{\rm S}.
\end{equation}

The magnification of the image is factored into tangential $\mu_t=( \gamma/\alpha)^{-1}$ and radial $\mu_r=( {\rm d}\gamma/ {\rm d}\alpha)^{-1}$ components, $\mu=\mu_t\mu_r$,
\begin{align}
\mu _{t}=\left( 1-\dfrac{d_{{\rm LS}}}{d_{{\rm S}}}\frac{\Delta \varphi}{\alpha }\right) ^{-1},\quad
\mu_{r}=\left( 1-\dfrac{d_{{\rm LS}}}{d_{{\rm S}}}\frac{{\rm d}\Delta\varphi}{{\rm d}\alpha }\right) ^{-1},
\end{align}
and expressed relative to corresponding exterior SdS ones, $\mu_{t{\rm ex}}$ and $\mu_{r{\rm ex}}$, as
\begin{align}
\mu_{t}^{-1}&=\mu_{t{\rm ex}}^{-1}+\frac{\epsilon^{2}}{\alpha^{2}}\left(1-\frac{\alpha^{2}}{\beta^{2}}\right)^{\frac{3}{2}},\label{ut}\\
\mu _{r}^{-1}&=\mu _{r{\rm ex}}^{-1}-\frac{\epsilon^{2}}{\alpha
^{2}}\left( 1+2\frac{\alpha ^{2}}{\beta^{2}}\right) \left( 1-
\frac{\alpha ^{2}}{\beta^{2}}\right) ^{\frac{1}{2}},\label{ur}\\
\mu _{t{\rm ex}}^{-1}&=1+\tilde{\eta}(\Lambda)-\frac{\epsilon^{2}}{\alpha^{2}},\quad\mu_{r{\rm ex}}^{-1}=1+\tilde{\eta}(\Lambda) +\frac{\epsilon^{2}}{\alpha ^{2}},\label{murtex}
\end{align}
where, for the sake of simplicity, we have set $\tilde{\eta}(\Lambda)=(d_{{\rm LS}}/d_{{\rm S}})\eta(\Lambda)$, and introduced the Einstein ring angle $\epsilon=\sqrt{4GMd_{\rm LS}/d_{\rm L}d_{\rm S}}$. The interior fluid adds a positive term $\propto$ $(1-\alpha^2/\beta^2)^{3/2}$ to $\mu _{t{\rm ex}}^{-1}$, so $\mu _{t}$ is systematically reduced (demagnified) relative to $\mu _{t{\rm ex}}$, consistent with weaker lensing from a distributed interior mass. Conversely, the interior subtracts a positive term $\propto$ $(1-\alpha^2/\beta^2)^{1/2}$ from $\mu _{r{\rm ex}}^{-1}$, so $\mu _{r}$ tends to be enhanced (magnified) compared with the exterior solution due to cumulative lensing effects inside the fluid sphere. A priori, the total magnification may be either diminished or enhanced. We express it in terms of the total exterior magnification $\mu _{{\rm ex}}$ as
\begin{align}
\mu^{-1}&=\mu_{{\rm ex}}^{-1}-3\frac{\epsilon^{2}}{\beta^{2}}\left[ 1+\tilde{\eta}(\Lambda) -\frac{\epsilon^{2}}{\beta^{2}}\frac{\zeta (\alpha)}{3}\right]\left( 1-\frac{\alpha ^{2}}{\beta^{2}}\right) ^{\frac{1}{2}},\label{mu}\\
\mu _{{\rm ex}}^{-1}&=\left(\mu _{t{\rm ex}}\mu _{r{\rm ex}}\right)^{-1}=\left[ 1+\tilde{\eta}(\Lambda) \right]^{2}-\frac{\epsilon^{4}}{\alpha^{4}},\label{muex}
\end{align}
where we have defined a positive function
\begin{equation}
\zeta (\alpha)=\frac{\beta^{4}}{\alpha ^{4}}\left\{ 1-\frac{\alpha^{2}}{\beta^{2}}+\left( 1+2\frac{\alpha ^{2}}{\beta^{2}}\right) \left[ 1-\left( 1-\frac{\alpha ^{2}}{\beta^{2}}\right) ^{\frac{3}{2}}\right] \right\}.
\end{equation}
Notably, increasing $\Lambda$ systematically raises $\tilde{\eta}(\Lambda)$~(\ref{eta}), leading to demagnification of tangential, radial~(\ref{murtex}), and total exterior magnifications~(\ref{muex}), consistent with previous analyses~\cite{Sereno2008,Zhao}. The interior correction enters with a minus sign in front of the $\Lambda$--dependent term~(\ref{mu}), yielding a slight magnifying effect that is suppressed by the small factor $(\epsilon/\beta)^2$ and thus negligible compared to the dominant exterior demagnification induced by $\Lambda$.

\section{Numerical Results and Discussion}
We now evaluate our results as function of the image angle $\alpha$ over the range $[0.9\beta,\beta]$ for two illustrative setups, adopting $r_{\rm E}= r_{\rm S}=3\,{\rm Gpc}$ and $\Lambda=10^{-52}\,{\rm m}^{-2}$:
\begin{itemize}
    \item Galaxy-scale: $M=5\times10^{11}M_{\odot }$ and $r_{\rm B}=25\,{\rm kpc}$.
    \item Cluster-scale: $M=10^{14}M_{\odot }$ and $r_{\rm B}=250\,{\rm kpc}$.
\end{itemize}
\begin{figure}[htbp!]
\centering

\includegraphics[width=1\linewidth]{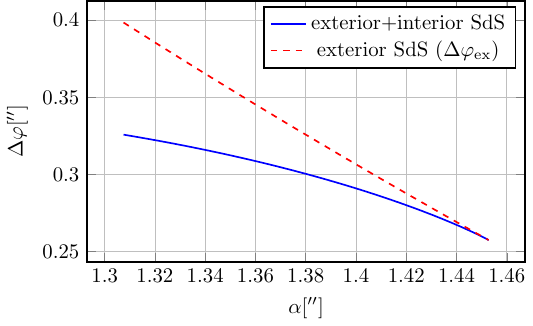}

\caption{Evolution of bending angle versus reception angle for a galaxy lens in matched exterior and interior SdS.}\label{defvsalpha_galaxy}
\end{figure}
\begin{figure}[htbp!]
\centering

\includegraphics[width=1\linewidth]{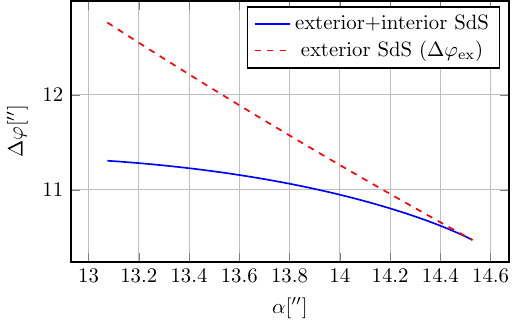}

\caption{Evolution of bending angle versus reception angle for a cluster lens in matched exterior and interior SdS.}\label{defvsalpha_cluster}
\end{figure}
\begin{figure}[htbp!]
\centering
\includegraphics[width=1\linewidth]{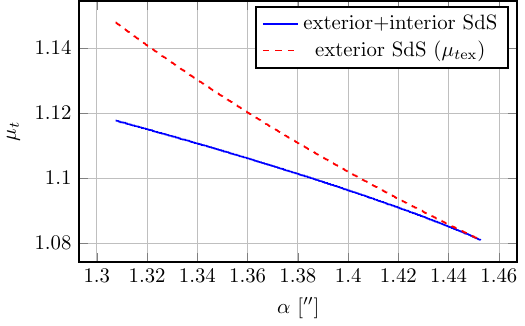}

\caption{Evolution of tangential magnification versus reception angle for a galaxy lens in matched exterior and interior SdS.}\label{magtvsalpha_galaxy}
\end{figure}
\begin{figure}[htbp!]
\centering

\includegraphics[width=1\linewidth]{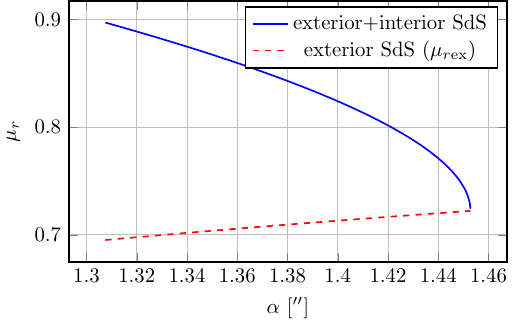}

\caption{Evolution of radial magnification versus reception angle for a galaxy lens in matched exterior and interior SdS.}\label{magrvsalpha_galaxy}
\end{figure}
\begin{figure}[htbp!]
\centering

\includegraphics[width=1\linewidth]{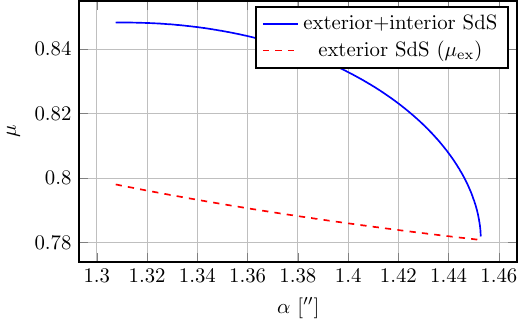}

\caption{Evolution of total magnification versus reception angle for a galaxy lens in matched exterior and interior SdS.}\label{magvsalpha_galaxy}
\end{figure}
\begin{figure}[htbp!]
\centering
\includegraphics[width=1\linewidth]{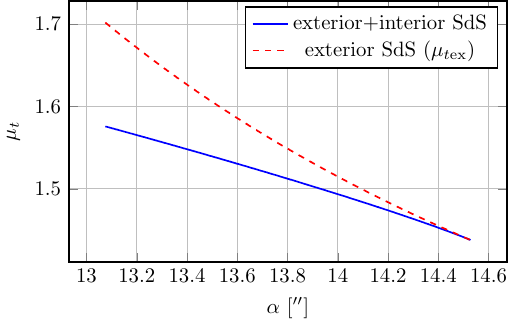}

\caption{Evolution of tangential magnification versus reception angle for a cluster lens in matched exterior and interior SdS.}\label{magtvsalpha_cluster}
\end{figure}
\begin{figure}[htbp!]
\centering

\includegraphics[width=1\linewidth]{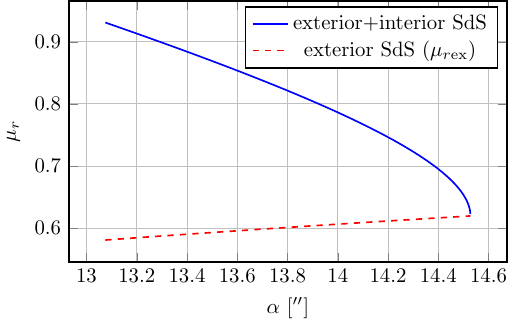}

\caption{Evolution of radial magnification versus reception angle for a cluster lens in matched exterior and interior SdS.}\label{magrvsalpha_cluster}
\end{figure}
\begin{figure}[htbp!]
\centering

\includegraphics[width=1\linewidth]{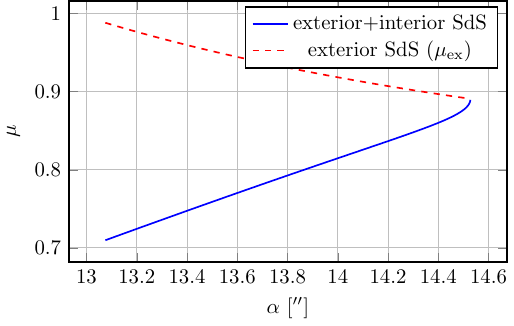}

\caption{Evolution of total magnification versus reception angle for a cluster lens in matched exterior and interior SdS.}\label{magvsalpha_cluster}
\end{figure}
The resulting $\Delta \varphi(\alpha)$, $\mu _{t}(\alpha)$, $\mu _{r}(\alpha)$, and $\mu(\alpha)$ curves are displayed in Figs.~\ref{defvsalpha_galaxy}, \ref{defvsalpha_cluster}, \ref{magtvsalpha_galaxy}, \ref{magrvsalpha_galaxy}, \ref{magvsalpha_galaxy}, \ref{magtvsalpha_cluster}, \ref{magrvsalpha_cluster}, and \ref{magvsalpha_cluster}, then compared against the pure exterior SdS predictions to isolate the impact of the interior structure.
The numerical trends confirm the analytic structure of Eqs.~(\ref{deftotal}), (\ref{ut}) and (\ref{ur}) for both galaxy and cluster scales: the interior terms systematically drive $\Delta \varphi\leq\Delta \varphi_{\rm ex}$ (Figs.~\ref{defvsalpha_galaxy} and \ref{defvsalpha_cluster}), reflecting the reduced curvature inside the fluid sphere relative to the vacuum exterior; $\mu_t\leq\mu_{t{\rm ex}}$ (Figs~\ref{magtvsalpha_galaxy} and \ref{magtvsalpha_cluster}), owing to the weaker angular focusing of a distributed mass compared to a point-like lens; and $\mu_r\geq\mu_{r{\rm ex}}$ (Figs.~\ref{magrvsalpha_galaxy} and \ref{magrvsalpha_cluster}), arising from the cumulative deflection accumulated along the interior path. The sign of the net correction to $\mu$ in Eq.~(\ref{mu}) is controlled by the compactness parameter $\epsilon^2/\beta^2(\propto\delta_{\rm B})$, leading to a net but slight magnification for the galaxy-scale (Fig.~\ref{magvsalpha_galaxy}), where the radial enhancement dominates, and a demagnification for the cluster case (Fig.~\ref{magvsalpha_cluster}), where tangential suppression prevails: for small $\epsilon^2/\beta^2$ (galaxy parameters) the leading $\mathcal{O}(\epsilon^2/\beta^2)$ term dominates, yielding $\mu\geq\mu_{\rm ex}$, while for larger $\epsilon^2/\beta^2$ (cluster parameters) the higher-order contribution flips the sign, producing $\mu\leq\mu_{\rm ex}$.

\section{Summary}
We have explored gravitational lensing within a matched Kottler spacetime, modeling the lens as a uniform-density fluid sphere smoothly joined to the exterior Kottler solution. Analytic expressions for the bending angle of light rays crossing both interior and exterior regions were derived, isolating distinctive corrections induced by the interior mass distribution compared to the vacuum SdS case. We demonstrated that the inclusion of interior structure systematically weakens the bending angle and tangential magnification, while enhancing the radial magnification, yielding scale-dependent signatures in total magnification. These analytic insights were corroborated by numerical examples, demonstrating that galaxy-scale lenses exhibit a slight net magnification, whereas cluster-scale lenses yield a net demagnification. This difference arises from lens compactness: more compact galaxy lenses produce radial magnification boosts, whereas diffuse clusters have dominant tangential demagnification. Additionally, the interior correction carries a small negative $\Lambda$--dependent contribution to the bending, consistent with its role in the vacuum SdS case. At the level of magnifications, $\Lambda$ induces a dominant exterior demagnification, while the interior term provides only a negligible magnifying effect suppressed by $(\epsilon/\beta)^2$. These results highlight the critical influence of interior structure on lensing predictions and motivate extensions to more realistic density profiles \cite{Navarro,Gabbanelli} and dynamical backgrounds \cite{Schu1,Guen1,Guen2,Guen3,Guen4}, which would clarify how robust these corrections to the deflection angle and magnification signatures are and whether they can serve as discriminants in forthcoming strong-lensing surveys. In Ref.~\cite{Guen5}, a companion study addresses the time delay in the same framework, offering a complementary probe of interior structure directly relevant to strong-lensing cosmography.

\bibliography{references}

\end{document}